\documentclass[doublecol]{epl2} 
\usepackage{amssymb}
\usepackage{amsmath}

\title{Statistical relativistic temperature transformation for ideal gas of bradyons, luxons and tachyons}
\shorttitle{Temperature transformation for bradyons, luxons and tachyons} 

\author{Felipe Asenjo \and Cristian A. Far\'\i as \and Pablo S. Moya}

\institute{
  Departamento de F\'isica, Facultad de
  Ciencias, Universidad de Chile - Las Palmeras 3425, \~Nu\~noa, Casilla
  653,\\ Santiago, Chile.
      }
      

\abstract{
Starting from a microcanonical statistical approach and special
relativity, the relativistic transformations for temperature and
pressure for an ideal gas of bradyons, luxons or tachyons is
found.  These transformations are in agreement with the three laws of
thermodynamic and our temperature transformation is the same as
Ott's. Besides, it is shown that the thermodynamic $dS$ element is
Lorentz-invariant.
}

\begin{document}

\maketitle

\section{Introduction}
The relativistic transformation of temperature is a problem which has
been controvertial for almost a century. There has been many proposals,
starting by pure classical
thermodynamics~\cite{Einstein,planck,tolman,Ott,landsvarianza,newburgh} until
classical and quantum statistical mechanics~\cite{bors,impos,kania1,kania2,cubero}. Starting from different
postulates, each one of this works had tried to establish how the 
different thermodynamics quantities change under Lorentz
transformations,
but they have obtained incompatible results. For example, in
Refs.~\cite{kampen,yuen} a review of different formalisms is done. In
particular, in Ref.~\cite{kampen} it is established that the different
formalisms are mathematically equivalent to each other, because there
is a one to one correspondence 
between the quantities defined in every formalism.

The idea to generalize the statistical mechanics to relativistic
systems dates from J\"uttner~\cite{juttner,juttner2}, who proposed a relativistic form of
Maxwell-Boltzmann velocities distribution. Other attempts to get the
correct relativistic distribution function that fits correctly
experimental data have been recently done. For example, in
Refs.~\cite{kania1,kania2} a new mathematical formalism was created in
order to develope a non-extensive relativistic statistical mechanics under a
canonical ensemble, which works to fit data of cosmic rays. Recently, it has been shown through
numerical simulation that J\"uttner's distribution function is the
distribution in special relativity that produces the best fit for a dilute gas of two components mixture with collisions
in one dimension~\cite{cubero}. 

In other hand, other works conduct to others distribution functions
than J\"uttner one. For instance, in Ref.~\cite{lapenta}, authors perfom
numerical simulations of electrons accelerated to relativistic
energies due to its interaction with waves generated by longitudinal
streaming plasma instabilities. They found an equilibrium
distribution which present power-law tails at high energies. Although
Refs.~\cite{cubero,lapenta} consider different systems, both show that
the old problems of transformation of temperature and pressure, and
the form of distribution function in theoretical relativistic
statistical mechanics appears in numerical simulation.  

However, in many of these works the temperature transformation are
assumed and not derived from the theory itself. This happen because in
canonical ensembles the temperature $\beta=T^{-1}$ is a system
variable (we set Boltzmann constant $k_B=1$). Therefore, it is not easy
to find a temperature transformation between two inertial frames
moving with relative velocities by direct calculation. 

To overcome this problem, in this article we calculate the temperature
in the microcanonical ensemble of a relativistic ideal gas of
bradyons, luxons or tachyons. In this ensemble the intensive quantities are not
variables and it is possible to find the temperature only by taking
derivatives. Thus, the calculations are
simpler than in a canonical ensemble because we only need to fix the energy of these
particles. In addition, according to Gibbs' postulate, the results
should be independent from the ensemble used to calculate it. This
postulate allow us to obtain a result that is equivalent to the one
obtained in any other ensemble~\cite{gibbs}. 

We are extending the old problem of how the temperature of
bradyons transform in different frames to luxons and tachyons. The reason to
include tachyons under this study is the wide range of relativistic
systems in which they can be included. They play an important role in
recent developments in inflationary cosmological
models~\cite{balart,frey,xiong}, string theory black holes
models~\cite{atish,rama} and there are, even, proposed procedures to
measure tachyonic states~\cite{chiao}.

To find the temperature transformation we first derive the
microcanonical entropy of the systems. Then we calculate the
temperature in a thermodynamic way showing how it
transforms. Futhermore, we show that the entropy thermodinamic $dS$
element is Lorentz invariant for each particle specie.

\section{Entropy calculation}

Consider an ideal gas (of bradyons, luxons or
tachyons) which is at rest in a inertial
frame $I$. Let us suppose other inertial frame $I'$ moving with constant velocity
$\mathbf{w}=w\hat x$ respective to $I$. 
Setting $c=1$, we choose the magnitude $w\leq1$ if the particles of the
systems are bradyons or luxons, and $w>1$ if the particle system are
tachyons. 

A bradyon is a particle with rest mass $m$
which moves slower than speed of light. Its dispersion relation is
given by  
\begin{equation}
  \label{erel}
  p_\mu p^\mu= m^2\, ,
\end{equation}
where $p_\mu=(\epsilon,\mathbf{p})$ is the 4-momentum of the particle with energy $\epsilon$
and momentum $\mathbf{p}$. We use the signature $(+,-,-,-)$ for our
calculations.

A luxon is a particle with null mass which moves
at the speed of light. Its dispersion relation has the form
 \begin{equation}
   \label{erell}
   p_\mu p^\mu=0\, . \end{equation}

Finally, a tachyon is a particle with imaginary mass $M=im$ (with $m$
a real quantity) which moves faster than speed of
light~\cite{feinberg,eve,mariwalla,maccarrone,feinberg2,kowa,antippa}.
Its dispersion relation is
\begin{equation}
  \label{erel2}
  p_\mu p^\mu=-m^2\, .
\end{equation}

We calculate the number of states $\Omega$ using the microcanonical
ensemble. The three-vector phase-space $d^3\mathbf{q}d^3\mathbf{p}$
is Lorentz invariant for bradyons, luxons and tachyons ~\cite{kowa}.

We consider an ideal gas of bradyons, luxons or tachyons,
consisting of $N$ particles ($N\gg1$) contained in a volume $V$. The Hamiltonian of $N$ bradyons is
\begin{equation}
H(p_i)=\sum_{i=1}^N\sqrt{|\mathbf{p}_i|^2+m^2}\, ,
\end{equation}
where $|\mathbf{p}_i|=(p_{x,i}^2+p_{y,i}^2+p_{z_i}^2)^{1/2}$. The Hamiltonian for $N$ luxons is
\begin{equation}
H(p_i)=\sum_{i=1}^N|\mathbf{p}_i|\, ,
\end{equation}
and the Hamiltonian for $N$ tachyons is
\begin{equation}
H(p_i)=\sum_{i=1}^N\sqrt{|\mathbf{p}_i|^2-m^2}\, .
\end{equation}

Setting $h=1$, the microcanonical number of states for each specie is given by 
\begin{eqnarray}
  \label{numestados}
  \Omega&=&\frac{1}{N!}\int_{E\leq H(p_i)\leq E+\Delta E} d^3\mathbf{q}_1\ldots d^3\mathbf{q}_Nd^3\mathbf{p}_1\ldots d^3\mathbf{p}_{N}\nonumber\\
&=&\frac{V^N}{N!}\int_{E\leq H(p_i)\leq E+\Delta E} d^3\mathbf{p}_1\ldots d^3\mathbf{p}_{N}\, .
\end{eqnarray}

For simplicity, we first calculate $\Sigma$ instead $\Omega$, where
\begin{equation}
\label{sigma}
\Sigma=\frac{V^N}{N!}\int_{H(p_i)\leq E} d^3\mathbf{p}_1\ldots d^3\mathbf{p}_{N}\, .
\end{equation}

The number of states in a energy interval can be calculated from $\Omega=(\partial
\Sigma/\partial E)\Delta E$. Thus, we must write the condition $H(p_i)\leq E$ in a
$3N$-dimensional momentum space. For photons $m=0$, and then 
\begin{equation}
H=\sum_i |{\bf p}_i|\leq E
\label{condmom0}
\end{equation}

Now, we seek for the condition for bradyons. Since no direction in space is preferred,
let us start supposing $n$ particles with the same momentum ${\bf p_0},$
and $N-n$ particles without momentum, with $n\leq N$. In this way, the
condition for the Hamiltonian is $H=n\left(|{\bf p_0}|^2+m^2\right)^{1/2}+(N-n)m\leq E$. Using this, we can obtain
$$\sum_i|{\bf p}_i|=n|{\bf p_0}|\leq\left((E-(N-n)m)^2-n^2m^2\right)^{1/2}$$
However, the factor $(E-(N-n)m)^2-n^2m^2=(E-Nm)(E-Nm+2nm)\leq E^2-N^2 m^2$, and then, it is fulfilled 
\begin{equation}
\sum_i|{\bf p}_i|\leq \left(E^2-N^2m^2\right)^{1/2}
\label{condmom}
\end{equation}
even when $n=N$.

Now, we should study what happen when we have different momenta for
each particle. An illustrative example is the next case. If we have
$N-1$ particles with the same momentum and one particle with a
different momentum, its sum of the norm of all momenta will be always
less than the sum of momenta in Eq.~(\ref{condmom}) owing to the
particles obey the condition $H\leq E$. Following this example, the
different cases of momentum of each particle will produced a sum which
will be less than Eq.~(\ref{condmom}). So, the condition Eq.~(\ref{condmom})
is always valid for bradyons.

Using an analogue argument we can obtain the condition for momentum space for tachyons
\begin{equation}
\sum_i|{\bf p}_i|\leq \left(E^2+N^2m^2\right)^{1/2}
\label{condmom2}
\end{equation}
which is fulfilled always.

All these conditions can be easily written for the momentum
components. Thus, the sum will go from 1 to $3N$. Written in that
form, they will represent a regular geometric body in $3N$ dimensions,
which would be a sphere in the case of classical ideal gas. Then, the
problem of calculate the integral Eq.~(\ref{sigma}) is reduced to find the
volume of this regular geometric body. Following the procedure
described in Ref.~\cite{greiner}, we obtain the number of states for
bradyons as  
\begin{equation}
  \label{numestadosb}
  \Omega=\frac{V^N}{N!}\left(2\sqrt 3\right)^{3N}\frac{\left(E^2- N^2m^2\right)^{3N/2}}{(3N)!}\, ,
\end{equation}
the number of states for luxons as
\begin{equation}
  \label{numestadosl}
  \Omega=\frac{V^N}{N!}\left(2\sqrt 3\right)^{3N}\frac{E^{3N}}{(3N)!}\, ,
\end{equation}
and the number of states of tachyons as
\begin{equation}
  \label{numestadost}
  \Omega=\frac{V^N}{N!}\left(2\sqrt 3\right)^{3N}\frac{\left(E^2+ N^2m^2\right)^{3N/2}}{(3N)!}\, .
\end{equation}

It is straigthforward to obtain the entropy as $S=\ln \Omega$ in a
microcanonical ensemble. For bradyons the entropy is
\begin{equation}
  \label{entropiab}
  S=N\ln\left[\frac{V(E^2- N^2m^2)^{3/2}}{27N^4}\right]+3N\ln \left[2\sqrt 3 e^{4/3}\right]\, .
\end{equation}

In the same way, the entropy for luxons is
\begin{equation}
  \label{entropial}
  S=N\ln\left[\frac{VE^3}{27N^4}\right]+3N\ln \left[2\sqrt 3 e^{4/3}\right]\, ,
\end{equation}
and the entropy for tachyons
\begin{equation}
  \label{entropiat}
  S=N\ln\left[\frac{V(E^2+ N^2m^2)^{3/2}}{27N^4}\right]+3N\ln \left[2\sqrt 3 e^{4/3}\right]\, .
\end{equation}

\section{Temperature transformation}

To find the relation between the
temperature of the system in the $I$ frame and the temperature in the
$I'$ frame we need to find how to calculate the number of
states in $I'$. According to Liouville theorem~\cite{misner} the 
dimensional phase space $d^3\mathbf{p}'d^3\mathbf{q}' = d^3\mathbf{p}d^3\mathbf{q}$ is Lorentz
invariant. Using this, the number of
states $\Omega'$ in the $I'$ frame can be written using the phase space of $I$ frame
\begin{multline}
\label{volps}
N!\, \Omega= \int_Id^3\mathbf{p}d^3\mathbf{q} \\\to N!\, \Omega'= \int_{I'}d^3\mathbf{p}'d^3\mathbf{q}' = \int_{I'}d^3\mathbf{p}d^3\mathbf{q}\, ,
\end{multline}
where the $I'$ subindex means that now the integration is for all
$\mathbf{p'}_j$ that satisfy $\sum_{i=1}^N |\mathbf{p'}_i|\leq ({E'^2-N^2
  m'^2})^{1/2}$ for bradyons, $\sum_{i=1}^N|\mathbf{p'}_i|\leq E'$ for
luxons and $\sum_{i=1}^N |\mathbf{p'}_i|\leq ({E'^2+N^2 m'^2})^{1/2}$ for
tachyons in the $I'$ frame.

Due to Eq.~(\ref{volps}), the entropy $S'$ calculated in the $I'$
frame has the same form of the entropy $S$ of Eq.~(\ref{entropiab}), Eq.~(\ref{entropial}), and
Eq.~(\ref{entropiat}), but
changing the energy $E$ by the energy $E'$, and the volume $V$ by the volume $V'$.

For bradyons and luxons the energy transforms as $E'=\gamma E$, the momentum
transforms as $p'=\gamma p$ and the
volume transform as $V=\gamma V'$ since the relative
movement is in one dimension. The relativistic factor is $\gamma=(1-w^2)^{-1/2}$ with $w\leq
1$. For this particles we are considering positive energies. 

In the case of tachyons, the energy and momentum transformations are $E'=\zeta E$ and
$p'=\zeta p$ respectively, where $\zeta=(w^2-1)^{-1/2}$ with
$w>1$~\cite{feinberg,maccarrone}. For simplicity, we consider the
positive momentum tachyons. Similarly, the volume transformation for tachyons is $V=\zeta V'$~\cite{eve}. Note
that if in the $I$ frame the energy, the momentum and the volume of tachyons are real
quantities, then in $I'$ frame these quantities are still real. 

The above energy, momentum and volume transformations are one of
the multiple transformations that can be constructed for a Lorentz
invariant tachyon-theory~\cite{eve,maccarrone,mariwalla,feinberg2}. Although the present
analysis can be done with other transformations, the election of the above one gives back
the usual and simpler energy and momentum relations for
tachyons~\cite{eve}. They ensure that the tachyon three-vector
phase-space $d^3{\bf q}d^3{\bf p}$ is invariant.

In order to obtain the temperature, we calculate the thermodynamical variation of
entropy. The variation is $dS=dE/T+(P/T)dV$, where the temperature $T$
and the pressure $P$ are defined by~\cite{greiner}
\begin{equation}
  \label{entrp11}
  \frac{1}{T}=\left(\frac{\partial S}{\partial E}\right)_V\, \quad,\quad \frac{P}{T}=\left(\frac{\partial S}{\partial V}\right)_E .
\end{equation}

In this way, the calculation of the temperature for bradyons, from
Eq.~(\ref{entrp11}), is
\begin{equation}
  \label{tempb}
  \frac 1 T =\frac{3NE}{E^2- N^2m^2}\, .
\end{equation}

The temperature for luxons is
\begin{equation}
  \label{templ}
  \frac 1 T =\frac{3N}{E}\, ,
\end{equation}
and the temperature for tachyons is
\begin{equation}
  \label{temt}
  \frac 1 T =\frac{3NE}{E^2+ N^2m^2}\, .
\end{equation}

Likewise, we can calculate the pressure for bradyons, luxons and tachyons from
Eq.~(\ref{entrp11}). This is
\begin{equation}
  \label{pressure}
  \frac P T =\frac{N}{V}\, ,
\end{equation}
for the three species. It corresponds to the state equation for an ideal gas.

Calculation of the temperature $T'$ for bradyons, luxons or tachyons in the $I'
$ frame
 can be done using Eq.~(\ref{volps}) to
 evaluate the entropy $S'$. We can write Eq.~(\ref{entrp11}) for intensive quantities in the
 $I'$ frame. This allow us to express Eq.~(\ref{tempb}) for bradyons,
 Eq.~(\ref{templ}) for luxons, and Eq.~(\ref{temt}) for tachyons in
 $I'$. Thus, we obtain how the temperature $T'$ from $I'$ frame
 transforms to temperature $T$ in the $I$ frame for bradyon and luxon ideal gas under the transformations
 for energy and momentum previously established
\begin{equation}
  \label{transT}
  {T'}={\gamma T}\, ,
\end{equation}
and for tachyon gas
\begin{equation}
  \label{transTt}
  {T'}={\zeta T}\, .
\end{equation}

The transformations in Eq.~(\ref{transT}) and in Eq.~(\ref{transTt}) implies that the
temperature is not a Lorentz invariant. The temperature transformation
for bradyons and luxons (\ref{transT}) is in coincidence with Ott's
temperature transformation~\cite{Ott} and other previous
works~\cite{newburgh,bors,sutcliffe,moya}, and it is in disagreement
with Planck's formalism~\cite{Einstein,planck,tolman,impos,kania2,kowa}. This means
that a moving gas of bradyons or luxons appears hotter. The temperature transformation for
tachyons~(\ref{transTt}) is derived for the first time in the
knowledge of the authors. 

The difference between our approach and other approaches is the
definition of temperature. We emphasize that temperature is defined in
a thermodynamic and statistical form by Eq.~(\ref{entrp11}). Thus, the
correct definition of Eq.~(\ref{entrp11}) leads naturally to the above
correct temperature transformations. 

We can do the same analysis for the pressure $P'$ in the $I'$ frame
using the transformation for the energy and the momentum. In the same
way, according to Liouville theorem and using Eq.~(\ref{entrp11}) and
Eq.~(\ref{pressure}) for $I'$, we can get the transformation of
pressure $P'$ from $I'$ frame to pressure $P$ in the $I$ frame for
bradyons and luxons as
\begin{equation}
  \label{transp}
  {P'}={\gamma^2 P}\, .
\end{equation}

Similarly, for tachyons, the pressure transformation is
\begin{equation}
  \label{transpt}
  {P'}={\zeta^2 P}\, .
\end{equation}

We also see that pressure is not Lorentz invariant. The result for bradyons and luxons coincide with the
same one found previously in Refs.~\cite{moya,sutcliffe}. This
transformation for pressure goes in contradiction with some previous
works~\cite{Einstein,planck,tolman,impos,kowa}. The tachyon
pressure transformation is derived for first time. Our thermodynamic and
statistical definition for pressure transformations~(\ref{transp})
and~(\ref{transpt}) preserves the properties of ideal gases. Thus, the
temperature and pressure transformation are necessary to get 
an ideal gas of any of these particles in both frames. After taking
into account Eq.~(\ref{transT}) and Eq.~(\ref{transp}), we can write
Eq.~(\ref{pressure}) in the $I'$ frame as
\begin{equation}
\label{gasI'}
P'\, V' = N\, T'\,.
\end{equation}

From this we can conclude that in every inertial frame an ideal
gas behaves like an ideal gas under Lorentz transformations as one would
expect due to special relativity principle. 

In the same token, the transformations of Eq.~(\ref{transT}),
Eq.~(\ref{transTt}), Eq.~(\ref{transp}) and Eq.~(\ref{transpt})
for intensive quantities $T$ and $p$, for bradyons, luxons
or tachyons satisfy
\begin{equation}
  \label{dS/dE}
   \left(\frac{\partial S'}{\partial E'}\right)_{V'} dE' = \left(\frac{\partial S}{\partial E}\right)_{V} dE\,,
\end{equation}
and
\begin{equation}
  \label{dS/dV}
   \left(\frac{\partial S'}{\partial V'}\right)_{E'} dV' = \left(\frac{\partial S}{\partial V}\right)_{E} dV\,.
\end{equation}

Therefore, the variation of the entropy is the same in both frames, which means
\begin{equation}
  \label{dS}
  dS = dS'\,,
\end{equation}
for any of the three species, wich is according to all previous works.

Finally, it is easy to get from Eq.~(\ref{templ}) the correct energy for
an ideal gas of luxons as $E=3NT$. From Eq.~(\ref{tempb}) we can
obtain the correct non-relativistic energy for an ideal gas of
bradyons when $m\gg T$ 
\begin{equation}
  \label{Ebnorel}
 E\simeq\frac{3}{2}NT+Nm\, .
\end{equation}

For an ideal gas of tachyons, it is possible to obtain the energy in
the very-high temperature limit when $m\ll T$. From Eq.~(\ref{temt}) we obtain
\begin{equation}
  \label{Etnorel}
 E\simeq\frac{Nm^2}{3T}\, .
\end{equation}

From~(\ref{Etnorel}) we can see that the energy become null when the tachyon velocity and temperature goes to
infinite as expected~\cite{feinberg}.

\section{Conclusions}

We have shown a new path to get some known results of temperature
transformation for a gas of non-interacting particles. Our treatment is from statistical first principles, only
assuming the known space-time and
energy-momentum Lorentz transformation along with Liouville theorem and Gibss' postulate.

The temperature transformation for a classical ideal gas composed of
particles which moves slower, equal or faster than light was shown
explicitely. These transformations are the correct ones, at least for
microcanonical ensemble, because they were derived using 
only the known statistical properties for each particle. In addition,
Liouville theorem allow us to work in the microcanonical ensemble for
any inertial frame, so the transformations obtained preserve the form
of the first and second laws of thermodynamics in all inertial
frames. This is a difference with the usual relativistic
thermodynamics treatment, where the forms of the first and second laws
are choosen in a more arbitrary way.

An interesting consecuence of the transformations that we
found for temperature and pressure is that the equation of
state of an ideal gas is a Lorentz invariant. This is in agreement
with the first postulate of special relativity as one would expect.

For tachyons, Eq.~(\ref{Etnorel}) is correct in the high temperature
limit, when their velocity goes to infinite and their
energy $E\to 0$. However, when the relative speed between frames
$w$ goes to infinite, the temperature $T'$ goes to zero from
Eq.~(\ref{transTt}). This is due to a tachyon in $I$ frame is a bradyon
in a $I'$ frame which moves with speed greater than $1$ relative to the $I$
frame~\cite{antippa}. The behavior of the tachyon temperature transformation~(\ref{transTt}) is
equivalent  to temperature transformation of bradyons when $w\to
1$. This shows the duality bradyon-tachyon between frames moving at
relative speeds greater than light.

\acknowledgments

We thank to Dr. Gonzalo Guti\'errez, Dr. J. Alejandro Valdivia and
MSc. Andr\'es Gom\'ez for useful discussions and their enlightening
comments. 

F. A. is grateful to Programa MECE Educaci\'on Superior for a Doctoral
Fellowship, C. A. F. is grateful to CONICyT Master Fellowship and
P. S. M. is grateful to CONICyT Doctoral Fellowship.


\begin{thebibliography}{34}
\bibitem{gibbs} However, it is an open question if the Gibbs' postulate requires modification in special relativity.
\end{thebibliography}
\end{document}